\documentclass[aps, a4paper, twocolumn]{revtex4}
\usepackage{amsfonts, amssymb, latexsym, revsymb, graphicx, verbatim}

\begin{document}
\title{An area law for the entropy of low-energy states}
\author{Llu\'\i s Masanes}
\affiliation{ICFO-Institut de Ciencies Fotoniques, Mediterranean Technology Park, 08860 Castelldefels (Barcelona), Spain}

\begin{abstract}
It is often observed in the ground state of quantum lattice systems with local interactions that the entropy of a large region is proportional to its surface area. In some cases, this area law is corrected with a logarithmic factor. This contrasts with the fact that in almost all states of the Hilbert space, the entropy of a region is proportional to its volume. This paper shows that low-energy states have (at most) an area law with the logarithmic correction, provided two conditions hold: (i) the state has sufficient decay of correlations, (ii) the number of eigenstates with vanishing energy-density is not exponential in the volume. These two conditions are satisfied by many relevant systems. The central idea of the argument is that energy fluctuations inside a region can be observed by measuring the exterior and a superficial shell of the region.
\end{abstract}

\maketitle

\section{Introduction}

Entropy quantifies the uncertainty about the state of a physical system. A bipartite system in a pure state has zero entropy, but the reduced state of one subsystem may have positive entropy. This is due to quantum correlations between the two subsystems, the entanglement. In fact, this entropy quantifies the entanglement in the sense of quantum information theory \cite{S ent}.

\medskip In classical physics, the entropy of a region inside a spatially-extended system at finite temperature is proportional to the volume of the region---entropy is an extensive quantity. At zero temperature, it is small and independent of the region. In quantum physics, at finite temperature, the entropy of a region is also proportional to the volume. But it has been observed in several models that, at zero temperature, the entropy of a region is proportional to its surface area \cite{80s, Srednicki, Audenaert, Vidal, Botero, Jin, Calabrese, Plenio, Casini}. In some models of critical free fermions the entropy scales as the area times the logarithm of the volume \cite{Wolf, Gioev}. This has been presented as a violation of the area law, although the dimensionality of the scaling is still that of the area. A celebrated proof shows that any one-dimensional system with finite-range interactions and an energy gap above the ground state obeys a strict area law \cite{1D}.

\medskip The original motivation for this problem is the analogy with black-hole physics, where the thermodynamic entropy is proportional to the surface area of the event horizon \cite{Bekenstein, 80s, Srednicki}. The second motivation is to guide the development of efficient methods for simulating quantum systems with classical computers. The number of parameters needed for specifying an arbitrary pure state of an $N$-partite system is exponential in $N$. If the state is not entangled, the number of parameters is proportional to $N$. Hence, there seems to be a correspondence between entanglement and complexity. In one spatial dimension, the relation between entropy and the complexity of simulating a system is well understood \cite{V,Schuch, Vidal}. The third motivation is to understand the kind of states that arise in quantum many-body systems with strong interactions. Almost all states in the Hilbert space obey a volume law for the entropy \cite{Hayden}. Hence, area laws tell a lot about the multipartite entanglement structure. At a finer level, the specific form of an area law tells additional information about the system: the logarithmic correction is a signature of criticality \cite{Audenaert, Vidal, Wolf, Gioev, Calabrese}; and the appearance of a negative constant is a signature of topological order \cite{topo}. For further overview of the topics around area laws see the review article cited \cite{review}.

\section{Results and summary} 

Consider an arbitrary hamiltonian $H$ with finite-range interactions in an $s$-dimensional lattice. The eigenstates have a well-defined global energy, but inside a region ${\cal X}$ of the lattice the energy may fluctuate. (The nomenclature of FIG.~1 is followed.) In Section III it is proven that these fluctuations can be observed by measuring the exterior of the region and a superficial shell inside the region, that is ${\bar{\cal X}} \cup {\cal S}$. In Section IV a condition is imposed to the ground state: if the operator $X$ has support on the region ${\cal R}$ which is separated from the support of the operator $Y$ by a distance $l$, then the connected correlation function decays at least as
\begin{equation}\label{eq 1}
	|\langle XY \rangle -\langle X\rangle \langle Y \rangle|
	\leq (l- \xi \ln\! |{\cal R}|)^{-s}\ ,
\end{equation}
where $\xi$ is a constant. This implies that energy fluctuations inside the region ${\cal X}$ cannot be observed in its bulk, namely ${\cal R}$. This provides a characterization for the approximate support of the global ground state inside the region ${\cal R}$. In Section V a condition on the density of states is assumed: if $H_{\cal X}$ is the subhamiltonian with all terms of $H$ whose support is fully contained in ${\cal X}$, then the number of eigenvalues lower than $e$ is bounded by
\begin{equation}\label{eq 2}
	\Omega_{\cal X} (e) \leq (\tau |{\cal X}|)^{\gamma (e-e_0) +\eta |\partial {\cal X}|}\ ,
\end{equation}
where $e_0$ is the lowest eigenvalue and $\tau, \gamma, \eta$ some constants independent of ${\cal X}$. This condition is only assumed for $e \sim |\partial {\cal X}|$. This implies an upper-bound on the dimension of the above-defined support subspace. This is used to bound the Von Newmann entropy for the reduction of the global ground state in the region ${\cal R}$
\begin{equation}\label{eq 3}
	S(\rho_{\cal R}) = 
	\mathrm{tr}\!\left( -\rho_{\cal R} \ln \rho_{\cal R} \right)
	\leq \mathrm{const}\, |\partial {\cal R}| \ln\! | {\cal R}|\ .
\end{equation}
Section VI contains a simpler proof for the area law (\ref{eq 3}) without assuming (\ref{eq 1}), but assuming (\ref{eq 2}) for all the range of $e$. In Section VII the above results for the ground state are generalized to other low-energy states (not necessarily eigenstates). Section VIII contains the conclusions.

\begin{figure}
\begin{center}
	\includegraphics[width=87mm]{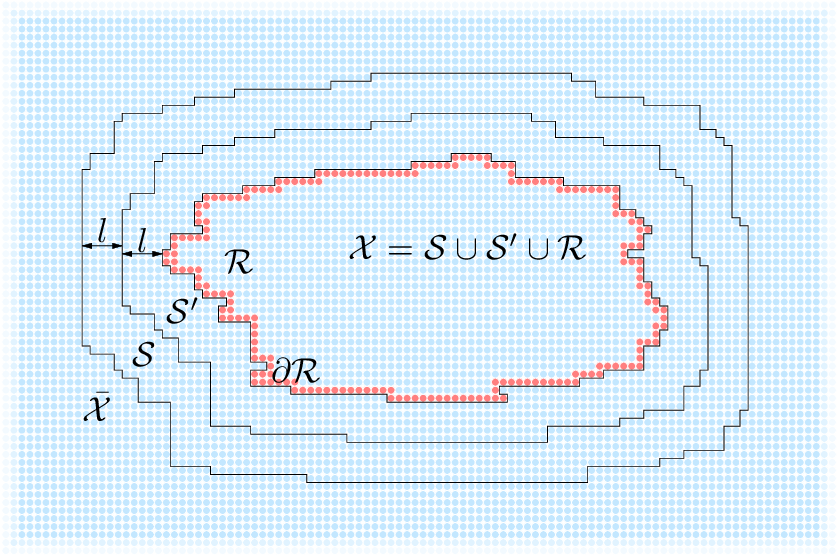}
	\caption{(Color online) ${\cal R}$ is the chosen region where the entropy is estimated; the sites belonging to its boundary $\partial {\cal R}$ are darker; ${\cal S}$ and ${\cal S}'$ are two superficial shells with thickness $l$ outside ${\cal R}$; ${\cal X} = {\cal S} \cup {\cal S}' \cup {\cal R}$ is the extended region; ${\bar{\cal X}}$ is the exterior of ${\cal X}$.}
	\label{fig 1}
\end{center}
\end{figure}

\section{Locality and energy fluctuations}

\subsection{Local interactions}

Consider a system with one particle at each site $x\in {\cal L}$ of a finite $s$-dimensional cubic lattice ${\cal L} \subset \mathbb{Z}^s$. The distance between two sites $x,y\in {\cal L}$ is defined by
\begin{equation}\label{distance}
  \mathrm{d}(x,y)= \max_{1\leq i \leq s} |x_i -y_i|\ .
\end{equation}
In the case of periodic boundary conditions or hybrids, this distance has to be modified with the appropriate identification of sites. Each particle $x\in {\cal L}$ has associated a Hilbert space with finite dimension $q$. 

\medskip The hamiltonian of the system can be written as
\begin{equation}\label{H}
	H= \sum_{x\in {\cal L}} K_x\ ,
\end{equation}
where each term $K_x$ can have nontrivial support on first neighbors ($y\in {\cal L}$ such that $\mathrm{d}(y,x) \leq 1$). There is a constant $J$ which bounds the operator norm of all terms $\|K_x \| \leq J$. (The operator norm of a matrix is equal to its largest singular value.) Translational symmetry is not assumed, so each term $K_x$ is arbitrary. The eigenstates and eigenvalues of $H$ are denoted by
\begin{equation}\label{eigenstates}
	H |\Psi_n\rangle = E_n |\Psi_n\rangle\ ,
\end{equation}
where the index $n=0,1,\ldots$ labels the eigenvalues in increasing oreder $E_n \leq E_{n+1}$. 

\medskip Note that any hamiltonian with finite-range interactions in a sufficiently regular lattice can be brought to the form of $H$, by coarse-graining the lattice. Quantum field theories with local interactions can also be brought to the form of $H$ by lattice regularization. In the case of bosons, a truncation in the local degrees of freedom is needed. In the case of fermions, a multi-dimensional Jordan-Wigner transformation \cite{VC} is needed.

\subsection{The Lieb-Robinson Bound}

The hamiltonian $H$ satisfies the premises for the Lieb-Robinson Bound \cite{LR, exp clustering}. Let $X,Y$ be two operators acting respectively on the regions ${\cal X}, {\cal Y} \subset {\cal L}$, with $\| X\|, \| Y\| \leq 1$. The distance between two regions is defined by 
\begin{equation}\label{distance}
	\mathrm{d}({\cal X}, {\cal Y}) = \min_{x\in {\cal X},\, y\in {\cal Y}} \mathrm{d}(x,y)\ .
\end{equation}
The time-evolution of an operator in the Heisenberg picture is $X(t)= e^{iHt} X e^{-iHt}$. The Lieb-Robinson Bound states that 
\begin{equation}\label{LR}
	\big\| [X(t),Y] \big\|
	\leq 2\, |{\cal X}|\,	\frac{(vt)^{\lfloor \mathrm{d}({\cal X}, {\cal Y})/2 \rfloor }}{\lfloor \mathrm{d}({\cal X}, {\cal Y})/2 \rfloor!}\ ,
\end{equation}
where $v=2 J 5^s$. When $vt \ll \mathrm{d}({\cal X}, {\cal Y})$ the two operators almost commute. In other words, the dynamics generated by $H$ does not allow for the propagation of signals at speed much larger than $v$. A simple proof of the Lieb-Robinson bound (\ref{LR}) is provided in Appendix C.

\subsection{Average for the energy fluctuations}

For any region ${\cal X} \subset {\cal L}$ and any integer $l \geq 5$ define the exterior, the boundary and the superficial shell as
\begin{eqnarray}\label{E}
  {\bar{\cal X}} &=& {\cal L}\backslash {\cal X}
	=\{ x\in {\cal L} : x\notin {\cal X} \}\ ,
	\\ \label{def dX}
	\partial {\cal X} &=& \{ x\in {\cal X} : \mathrm{d}(x, \bar{\cal X}) =1\} \ ,
	\\ \label{S2}
	{\cal S} &=& \{x\in {\cal X}: \mathrm{d}(x, {\bar{\cal X}})\leq l\}\ ,
\end{eqnarray}
respectively (see FIG.~1). The hamiltonian $H_{\cal X}$ is defined as the sum of all terms $K_x$ whose support is fully contained in ${\cal X}$. The eigenstates and eigenvalues of $H_{\cal X}$ are denoted by 
\begin{equation}\label{eigenstates}
	H_{\cal X} |\psi_n\rangle = e_n |\psi_n\rangle\ ,
\end{equation}
where the index $n=0,1,\ldots$ labels the eigenvalues in increasing order $e_n \leq e_{n+1}$. The sum of all terms $K_x$ which simultaneously act on both, ${\cal X}$ and ${\bar{\cal X}}$, is denoted \mbox{$H_1 = H- H_{\cal X} -H_{\bar{\cal X}}$}, and has norm $\| H_1 \| \leq J 3^s |\partial {\cal X}|$. The expectation of any operator $X$ with the ground state is denoted by $\langle X \rangle = \langle \Psi_0 |X| \Psi_0\rangle$. Without loss of generality it can be assumed that each $K_x$ is positive semi-definite, which implies
\begin{eqnarray*}
	&&\hspace{-10mm}	
	\langle H_{\cal X} \rangle + \langle H_{\bar{\cal X}} \rangle
	\\  &\leq& 
	\langle H_{\cal X} + H_1 + H_{\bar{\cal X}} \rangle
	\\  &\leq&
	\mathrm{tr} \big[(H_{\cal X} +H_1 +H_{\bar{\cal X}}) (|\psi_0\rangle\!\langle \psi_0 | \otimes
	\mathrm{tr}_{\cal X} |\Psi_0\rangle\!\langle \Psi_0 |) \big]
	\\   &\leq&
	e_0 + J\, 3^s\, |\partial {\cal X}| +\langle H_{\bar{\cal X}} \rangle \ ,
\end{eqnarray*}
and
\begin{equation}\label{e bound}
	e_0\ \leq\ \langle H_{\cal X} \rangle\ \leq\ 
	e_0 + J\, 3^s\, |\partial {\cal X}|\ .
\end{equation}
This can be sumarized as follows.

\medskip{\em The energy frustration of the global ground state $|\Psi_0\rangle$ in a region ${\cal X}$ is, at most, proportional to the boundary $\partial {\cal X}$.}

\subsection{Observation of energy fluctuations}

For any value of $e_\mathrm{cut}$ define the operator
\begin{equation}\label{Q}
	Q= \int_{-\infty}^{e_\mathrm{cut}} \hspace{-3mm} d\omega \int\! \frac{dt}{2\pi}\, 
	e^{-\frac{\sigma t^2}{2}}\, e^{i(E_0 -\omega)t}\, e^{iH_{\cal X} t}\, 
	e^{-i H t}\ ,
\end{equation}
where $\sigma = 10^4 v^2/ l$. The action of $Q$ onto the global ground state $|\Psi_0\rangle$ implements an approximate projection onto the subspace with energy lower than $e_\mathrm{cut}$ inside the region ${\cal X}$,
\begin{equation}\label{step}
	Q|\Psi_0\rangle = \left[ \sum_n 
	\int_{-\infty}^{e_\mathrm{cut} -e_n}
	\hspace{-5mm} d\omega\  
	\frac{e^{-\frac{\omega^2}{2\sigma}}}{\sqrt{2\pi \sigma}}	\ |\psi_n \rangle\!\langle \psi_n|	\right]\!
	|\Psi_0\rangle\ .
\end{equation}
This integral is the error function, which is a soft step function. In the limit where the softness parameter $\sigma$ tends to zero, the operator inside the square brackets becomes a projector. The operator $Q$ has non-trivial support on the whole lattice ${\cal L}$, but remarkably, it can be approximated by the operator
\begin{equation}\label{tilde Q}
	\tilde{Q} = \int_{-\infty}^{e_\mathrm{cut}} \hspace{-3mm} d\omega 
	\int\! \frac{dt}{2\pi}\, 
	e^{-\frac{\sigma t^2}{2}}\, e^{i(E_0 -\omega)}\, e^{iH_{{\cal S}} t}\, 
	e^{-i H_{{\bar{\cal X}}\cup {\cal S}} t}\ ,
\end{equation}
which has non-trivial support only in the region ${\bar{\cal X}} \cup {\cal S}$. More quantitatively, the bound 
\begin{equation}\label{Q tildeQ}
	\|Q -\tilde{Q}\| \leq |{\cal X}|^3 \, e^{-l}
\end{equation}
is proven in Lemma 1 (Appendix), using techniques similar to the ones in \cite{1D, exp clustering, spectrum cond}. The fact that $Q\approx \tilde{Q}$ is solely a consequence of the locality of interactions and can be understood as follows. According to the Lieb-Robinson bound (\ref{LR}), if $t<l/v$, any operator $Y$ with support on ${\cal X} \backslash {\cal S}$ evolves to an operator $Y(t)$ with approximate support on ${\cal X}$. Then $e^{-i H_{\cal X} t} Y(t)\, e^{i H_{\cal X} t} \approx Y$, or in other words, the unitary $e^{i H_{\cal X} t} e^{-i Ht}$ in (\ref{Q}) approximately acts like the identity inside ${\cal X}\backslash {\cal S}$, or in other words $e^{i H_{\cal X} t} e^{-i Ht} \approx e^{iH_{{\cal S}} t}	e^{-i H_{{\bar{\cal X}}\cup {\cal S}} t}$, which justifies the definition (\ref{tilde Q}).

\medskip The right-hand side of (\ref{tilde Q}) is an average of unitaries, therefore $\|\tilde{Q} \| \leq 1$. Then, the operators $|\tilde{Q}|$ and $(\mathbb{I}- |\tilde{Q}|)$ define a two-outcome generalized measurement on ${\bar{\cal X}}\cup {\cal S}$, which tells whether the energy inside ${\cal X}$ is below or above $e_\mathrm{cut}$, approximately.

\medskip Everything shown in this section for the ground state generalizes to all eigenstates. The action of $Q$  onto $|\Psi_n \rangle$ is
\begin{equation}\label{step n}
	Q|\Psi_n\rangle = \left[ \sum_n 
	\int_{-\infty}^{e'_\mathrm{cut} -e_n}
	\hspace{-5mm} d\omega\  
	\frac{e^{-\frac{\omega^2}{2\sigma}}}{\sqrt{2\pi \sigma}}	\ |\psi_n \rangle\!\langle \psi_n|	\right]\!
	|\Psi_n\rangle\ ,
\end{equation}
where $e'_\mathrm{cut} = e_\mathrm{cut} +E_n -E_0$. Summarizing, for each eigenstate $|\Psi_n \rangle$ there is an operator $\tilde{Q}$ which approximately projects onto the subspace with energy $e_n \leq e_\mathrm{cut}$ inside the region ${\cal X}$, by only acting on the exterior and the shell ${\bar{\cal X}} \cup {\cal S}$. The degree of approximation increases with $l$, the width of ${\cal S}$. The larger $l$ is, the closer $Q$ and $\tilde{Q}$ are, and the smaller the softness parameter $\sigma$ is. 

\medskip{\em The energy fluctuations of an eigenstate $| \Psi_n \rangle$ inside a region ${\cal X}$ can be observed by measuring the exterior and a superficial shell inside the region, that is ${\bar{\cal X}}\cup {\cal S}$} (see FIG.~1).

\section{Support of the ground state inside a region}

\subsection{Decay of correlations} 

It is usually the case that, when the system is in the ground state, the correlation between two observables acting on different sites decrease with the distance between the sites. Let $\Gamma$ be a function which upper-bounds the connected correlation function of any pair of operators $X,Y$ acting respectively on the disjoint regions ${\cal X}, {\cal Y} \subset {\cal L}$, with $|{\cal X}| \leq  |{\cal Y}|$ and $\| X\|, \| Y\| \leq 1$, 
\begin{equation}\label{clustering}
	\big| \langle XY\rangle -\langle X\rangle \langle Y\rangle \big|
	\leq \Gamma\big( \mathrm{d}({\cal X}, {\cal Y}), |{\cal X}| \big)\ .
\end{equation}
For the argument of this paper, both, the decay with the distance $\mathrm{d}({\cal X}, {\cal Y})$ and the scaling with size of the support of the operators $|{\cal X}|$, are relevant. It is shown in \cite{exp clustering} that any hamiltonian $H$ with an energy gap above the ground state $\Delta = E_1 - E_0 >0$ has
\begin{equation}\label{exponential clustering}
	\Gamma(l, |{\cal X}|)
	= c_1 |{\cal X}|\, e^{-l/\xi}\ ,
\end{equation}
with correlation length $\xi = 10 v/\Delta$. To prove the area law for the entropy the following condition is needed.

\medskip \noindent {\bf Assumption 1 } The correlation functions for the ground state decay at least as
\begin{equation}\label{poly clustering}
	\Gamma(l, |{\cal X}|)
	= \frac{c_1}{(l- \xi \ln |{\cal X}|)^{\nu}}\ ,
\end{equation}
where $c_1 ,\xi$ and $\nu >s$ are constants. 

\medskip\noindent Note that both, (\ref{exponential clustering}) and (\ref{poly clustering}), have the same relative scaling of $l$ and $|{\cal X}|$, but assumption (\ref{poly clustering}) is weaker than (\ref{exponential clustering}). Although the decay (\ref{poly clustering}) is polynomial in $l$, it is not the correlation function of a critical hamiltonian, where one expects $\Gamma \sim (|{\cal X}|^{1/s} /l)^{\nu}$. Unfortunately, the argument of this paper does not give an area law with such scaling in $|{\cal X}|$. 

\subsection{Energy fluctuations inside a region cannot be observed in its bulk}

For any region ${\cal R} \subset {\cal L}$ and any integer $l\geq 5$ define the extended region as
\begin{equation}\label{def X}
	{\cal X} = 
	\{x\in {\cal L}: \mathrm{d}(x,{\cal R})\leq 2 l\}\ , 
\end{equation}
which redefines (\ref{E}), (\ref{def dX}) and (\ref{S2}) (see FIG.~1). The region ${\cal R}$ can be considered the bulk of ${\cal X}$. 

\medskip Suppose the existence of an operator $Z$ with support in ${\cal R}$ such that 
\[
	Z |\Psi_0\rangle \approx \hspace{-2mm}
	\sum_{n:\, e_n \leq e_\mathrm{cut}} \hspace{-2mm} 
	|\psi_n\rangle\! \langle\psi_n | \Psi_0\rangle\ .
\]
This operator acts onto the ground state in a similar way as $\tilde{Q}$ does, then the two operators are correlated 
\[
	\langle Z \tilde{Q} \rangle \approx \langle Z \rangle \approx \langle \tilde{Q} \rangle\ ,
\]
and their corresponding supports are separated by a distance $l$. For the right choice of $e_\mathrm{cut}$ and large enough $l$ the existence of $Z$ is in contradiction with Assumption 1, therefore

\medskip {\em The energy fluctuations of the global ground state inside a region ${\cal X}$ cannot be observed in the bulk of the region, that is ${\cal R}$.}

\medskip In the following subsection, a quantitative example of this fact is given.

\subsection{Characterization of the support}

In what follows, the assignation 
\begin{equation}\label{ass cut}
	e_\mathrm{cut}= 2 J 3^s |\partial {\cal X}| + e_0 + 20 v
\end{equation}
is assumed in the definitions of $Q$ and $\tilde{Q}$ (\ref{Q},\ref{tilde Q}). 

\medskip\noindent {\bf Definition of $P$ } For each eigenstate $|\psi_n\rangle$ of $H_{\cal X}$ with $e_n \leq e_\mathrm{cut} + 20 v$ consider the Schmidt decomposition \cite{S ent} $|\psi_n \rangle = \sum_i \mu^i_n |\alpha^i_n \rangle \otimes |\beta^i_n \rangle$ with respect to the partition $|\alpha^i_n \rangle \in {\cal H}_{\cal R}$ and $|\beta^i_n \rangle \in {\cal H}_{{\cal X}\backslash {\cal R}}$. Define $P$ as the projector onto the subspace of ${\cal H}_{\cal R}$ generated by all vectors $|\alpha^i_n \rangle$ defined above, symbolically
\begin{equation}\label{P}
	P = \mathrm{supp}_{\cal R}
	\{ |\psi_n\rangle : e_n \leq 2 J 3^s |\partial {\cal X}| + e_0 + 40 v\} \ .
\end{equation}

\medskip\noindent Let $P^\bot = \mathbb{I}- P$ be the projector onto the complementary subspace. Lemma 3 (Appendix) shows that the assignation (\ref{ass cut}) implies
\begin{eqnarray}\label{tQ b}
	\langle \tilde{Q}\rangle &\geq& \frac{1}{2} - 
	2 |{\cal X}|^3 e^{-l}\ ,
\\ \label{b}
	\langle P^\bot \tilde{Q} \rangle 	&\leq&  
	2 |{\cal X}|^3 e^{-l}\ .
\end{eqnarray}
Recalling that the respective supports of $P^\bot$ and $\tilde{Q}$ are separated by a distance $l$, one can invoke the decay of correlations (\ref{clustering}) without specifying the function $\Gamma$,
\begin{equation}\label{QPGamma b}
	\langle P^\bot \rangle \langle \tilde{Q} \rangle -
	\langle P^\bot \tilde{Q} \rangle \leq \Gamma (l,|{\cal R}|) \ .
\end{equation}
The combination of (\ref{tQ b}), (\ref{b}) and (\ref{QPGamma b}) gives 
\begin{equation}\label{bound P}
	\langle P \rangle \ \geq\  
	1-4\, \Gamma (l,|{\cal R}|)\ ,
\end{equation}
for sufficiently large $l$, where $1/2 \geq \Gamma (l,|{\cal R}|) \geq 6|{\cal X}|^3 e^{-l}$ holds. Concluding, the support of the global ground state inside ${\cal R}$ is contained in the subspace characterized by $P$, up to some small weight (\ref{bound P}). 

\subsection{A renormalization group scheme} 

The projector $P$ defined above allows for certifiably-generating a low-energy effective theory for $H$: the hamiltonian terms $K_x$ inside ${\cal R}$ can be renormalized as 
\begin{equation}\label{RG}
	K_x \stackrel{\mathrm{RG}}{\longrightarrow} 
	P K_x P\ . 
\end{equation}
The whole lattice can be divided in similar regions, and the transformation (\ref{RG}) performed in each of them. The fidelity between the effective and the original ground-states can be bounded with (\ref{bound P}), and increased by enlarging $l$. As explained in Section VI, one can also obtain arbitrarily-good fidelities for any low-energy state. 

\section{Entanglement in the ground state}

\subsection{Energy spectrum}

In the previous section, a subspace which approximately contains the support of the ground state inside a region has been characterized. In order to bound its dimension, an additional assumption is needed: if the boundary conditions of the hamiltonian are left open, the number of eigenstates with vanishing energy-density must not be exponential in the volume.

\medskip \noindent {\bf Assumption 2 } There are constants $c_2, \tau, \gamma, \eta$ such that, for any region ${\cal X}$ and energy 
\begin{equation}\label{e cond}
	e =  2J 3^s |\partial {\cal X}| +e_0 +40v\ , 
\end{equation}
the number of eigenvalues of $H_{\cal X}$ lower than $e$ satisfies
\begin{eqnarray}\nonumber
	\Omega_{\cal X} (e) &=& \max\{n: e_n \leq e\}
	\\ \label{spectrum cond} &\leq& 
	c_2 (\tau |{\cal X}|)
	^{\gamma (e-e_0) + \eta |\partial {\cal X}|}\ .
\end{eqnarray}

\medskip\noindent The area law is nontrivial when applied to regions ${\cal R}$ such that $|\partial{\cal R}| \ll |{\cal R}|$, or equivalently $|\partial{\cal X}| \ll |{\cal X}|$. In this case, the eigenstates with energy proportional to the boundary $|\partial {\cal X}|$ (\ref{e cond}) have vanishing energy density $e_n /|{\cal X}|$. According to \cite{spectrum cond}, Assumption 2 holds for many systems that have an energy gap above the ground state. There are known hamiltonians which violate Assumption~2 and have a gap, but when the boundary conditions are opened there appears a degeneracy for the ground state which is exponential in the volume \cite{spectrum cond}. Massive free bosons and fermions satisfy Assumption~2. Contrary, massless free frermions violate it as $\Omega \sim \exp{\!\sqrt{(e-e_0) |{\cal X}|^{1/s} } }$. 

\medskip The factor $(\tau |{\cal X}|)^{\gamma (e-e_0)}$ in (\ref{spectrum cond}) can be understood with the following example. Consider the hamiltonian 
\[
	H_{\cal X} =		
	\sum_{x\in {\cal X}} 
	\left[
	\begin{array}{cc}
		1 & 0 \\
		0 & 0
	\end{array}
	\right]_x\ ,
\]
where the subindex $x$ specifies in which site the matrix acts. The energy $e\in \{0,1,\ldots |{\cal X}| \}$ counts the number of local excitations, hence the degeneracy is the binomial of $|{\cal X}|$ over $e$, which can be upper-bounded by $|{\cal X}|^e$. The constant factor $(\tau |{\cal X}|)^{\eta |\partial {\cal X}|}$ in (\ref{spectrum cond}) is introduced because some hamiltonians with open boundary conditions have a degeneracy (or approximate degeneracy) which is exponential in the size of the boundary. 

\medskip Consider again the Schmidt decomposition of each eigenstate $|\psi_n\rangle$ with respect to the partition ${\cal R}$ and ${\cal X}\backslash {\cal R}$ (Definition of $P$). The dimension of the Hilbert space ${\cal H}_{{\cal X}\backslash {\cal R}}$ is $q^{|{\cal X}\backslash {\cal R}|}$, therefore the support of each $|\psi_n \rangle$ on ${\cal R}$ has at most dimension $q^{|{\cal X}\backslash {\cal R}|}$. This and Assumption 2 provide a bound for the rank of the projector $P$
\begin{equation}\label{rank P}
	\mathrm{rank} P \ \leq\ q^{|{\cal X}\backslash {\cal R}|}\, 
	c_2 (\tau |{\cal X}|)
	^{[|\partial {\cal X}| (\gamma 2J 3^s + \eta) + \gamma 40 v]}\ .
\end{equation}

\subsection{Entropy of an arbitrary region}

Consider a region ${\cal R} \subset {\cal L}$ being a completely arbitrary subset of the lattice. It not need to be convex, full-dimensional nor connected. For any site $x$
\[
\begin{array}{lll}
	|\{y\in {\cal L}: \mathrm{d}(y,x) \leq 2l \}| 
	&\leq& (5l)^s\ ,
	\\
	|\{y\in {\cal L}: \mathrm{d}(y,x) = 2l \}| 
	&\leq& 2s (5l)^{s-1}\ , 
\end{array}
\]
which imply
\begin{eqnarray}
	\nonumber
	|{\cal X}| &\leq& |{\cal R}| (5l)^s\ ,
	\\	\label{b dX}
	|\partial {\cal X}| &\leq& |\partial {\cal R}| 2s (5l)^{s-1}\ ,
	\\ \nonumber
	|{\cal X}\backslash {\cal R}| &\leq& |\partial {\cal R}| (5l)^s\ .
\end{eqnarray}
Let $\rho_{\cal R} = \mathrm{tr}\!_{{\cal L} \backslash {\cal R}} |\Psi_0 \rangle\!\langle \Psi_0 |$ be the reduction of the ground state in ${\cal R}$, and $\lambda_1 \geq \lambda_2 \geq \cdots$ its eigenvalues in decreasing order. Assumptions 1 and 2 imply (\ref{poly clustering}), (\ref{bound P}) and (\ref{rank P}), which impose the following constraints on the eigenvalues: for any integer $l\geq 5$,
\begin{eqnarray}\label{constr}
	\sum_{k=1}^{\Theta(l)} \lambda_k &\geq&
	1- \theta(l)\ ,
	\\ \nonumber
	\theta(l) &=& \frac{4\, c_1}{(l- \xi\ln |{\cal R}|)^{\nu}}\ ,
	\\ \nonumber
	\ln \Theta(l) &=& |\partial {\cal R}| 
	2s (5l)^{s-1}
	\big( \gamma 2 J 3^s + \eta \big)
	\ln\!\big[ \tau |{\cal R}| (5l)^s \big]  
	\\ \nonumber && 
	+\ |\partial {\cal R}| (5l)^s \ln q 
	+ {\cal O}\big(\ln\! |{\cal R}|\big)\ .\end{eqnarray}
Now one can find the probability distribution $\lambda_k$ which maximizes the entropy $(-\sum \lambda_k \ln \lambda_k)$ given the above constraints. This is done in Appendix B with the following result.

\medskip\noindent {\bf Result 1 } The entropy of the reduction of the ground state inside an arbitrary region ${\cal R}$ satisfies
\begin{eqnarray}\nonumber
	S(\rho_{\cal R}) &\leq& |\partial {\cal R}|
	(10\, \xi \ln\!|{\cal R}|)^s
	\left[\frac{s}{\xi}\big(\gamma J\, 3^s + \eta \big) + \ln q\right]  
	\\
	&&+\ {\cal O}\big[ |\partial {\cal R}|\,
	(\ln\!|{\cal R}|)^{s-1} \big]\ .
\end{eqnarray}

\subsection{Entropy of a cubic region}

Consider the case where the chosen region is a hypercube ${\cal R} = \{x\in {\cal L}: 0\leq x_i \leq L\}$. One can proceed as before, but the bounds analogous to (\ref{b dX}) are smaller, implying a smaller bound for the entropy. All this is worked out in Appendix B. 

\medskip\noindent {\bf Result 2 } The entropy of the reduction of the ground state inside an cubic region ${\cal R}$ satisfies
\begin{equation}\label{S cubic}
	S(\rho_{\cal R}) \leq |\partial {\cal R}|
	\ln\!|{\cal R}|
	\big(\gamma 2 J 3^s + \eta + 4\xi \ln q \big)  
	+ {\cal O}\big( |\partial {\cal R}| \big)\ .
\end{equation}

\medskip \noindent It is expected that the entropy of any full-dimensional convex region ${\cal R}$ obeys the same scaling (\ref{S cubic}).

\section{Simpler proof for the area law}

An area law can be easily proven without Assumption~1, if Assumption~2 is extended to all values of the energy $e$, not only the ones satisfying (\ref{e cond}). Let ${\cal R}$ be the region where the entropy is estimated, and $H_{\cal R}$ the sum of all terms of the hamiltonian (\ref{H}) which are fully contained in ${\cal R}$. Following the conventions of this paper, the eigenstates and eigenvalues are denoted by $H_{\cal R} |\psi_n \rangle = e_n |\psi_n \rangle$, where $e_0 \leq e_1 \leq \cdots$. The strong version of Assumption~2 tells that all the eigenvalues $e_n$ satisfy 
\begin{equation}\label{a1}
	n \leq c_2 (\tau |{\cal R}|)^{\gamma (e_n-e_0) +\eta |\partial {\cal R}|}\ .
\end{equation}
The global ground state can be written as
\begin{eqnarray}
	|\Psi_0 \rangle = \sum_k \sqrt{\mu_n}\, |\psi_n \rangle \otimes
	|\varphi_n \rangle\ ,
\end{eqnarray}
where the coefficients $\mu_n$ are non-negative and add up to one. It is shown in \cite{S ent} that the entanglement entropy of $|\Psi_0\rangle$ is upper-bounded by the entropy of the $\mu$-coefficients
\begin{equation}\label{S bound}
	S(\rho_{\cal R}) \leq -\sum_n \mu_n \ln \mu_n\ .
\end{equation}
Locality implies equation (\ref{e bound}), which can be written as
\begin{equation}\label{contraint}
	\sum_n \mu_n e_n \leq e_0 + J3^s |\partial {\cal R}|\ .
\end{equation}
Maximizing the right-hand side of (\ref{S bound}) over the probability distribution $\mu_n$ and the numbers $e_n$ subjected to the constrains (\ref{a1}) and (\ref{contraint}) gives
\begin{equation}
	S(\rho_{\cal R}) \leq 2J3^s \gamma\, 
	|\partial{\cal R}| \ln |{\cal R}|\ ,
\end{equation}
the area law. This calculation is made in Appendix D.

\section{Entanglement in excited states}

The entanglement properties of excited states have also been studied. In references \cite{DS1,DS2,DS3} the motivation was to study the robustness of the area law for the entropy of black holes. They show that in systems of free bosons, the entropy of some low-energy excited states scales at most like the area. In \cite{AFC} the entropy scaling in one-dimensional spin systems is analyzed. They show that some excited states have entropy proportional to the volume, but low-energy states obey an area law. All this work is for integrable systems. In what follows, we address the general case.

\medskip Sometimes, low-lying excited states $|\Psi_n \rangle$ have correlation functions similar to the ones of the ground state. The single-mode ansatz for excitations with momentum $k$ is 
\begin{equation}\label{sm}
	|\Psi^\mathrm{sm}_k \rangle \propto \sum_x e^{ix \cdot k} Z_x |\Psi_0\rangle\ ,
\end{equation}
where $Z_x$ is an operator acting on site $x$ such that $\langle Z_x \rangle =0$. If $X,Y$ have support on finite regions and the volume of the system $|{\cal L}|$ tends to infinite, then the correlation function (\ref{clustering}) for the state (\ref{sm}) is the same as for $|\Psi_0 \rangle$. The same happens to excited states containing a small number of single-mode excitations. Examples of single-mode excitations are: spin waves, free bosons and free fermions. In this section it is shown that such excited states obey an area law similar to the one for the ground state. Actually, this is done with a bit more generality.

\medskip Consider an arbitrary superposition of eigenstates with bounded energy
\begin{equation}\label{Phi}
	|\Phi \rangle = \hspace{-3mm}
	\sum_{\ \ n:\, E_n \leq E_\mathrm{m}} 
	\hspace{-2mm} \mu_n\, |\Psi_n\rangle \ .
\end{equation}
In this case, the correct assignation for $e_\mathrm{cut}$ in the definitions of $Q$, $\tilde{Q}$ and $P$  (\ref{Q}, \ref{tilde Q}, \ref{P}) is
\begin{equation}
	e_\mathrm{cut} = 2J3^s |\partial {\cal X}| + e_0 + 20v + E_\mathrm{m}- E_0\ .
\end{equation}
Applying Assumption 1 to the state (\ref{Phi}), the arguments follow exactly as for the ground state. Repeating the calculation of the entropy for a cubic region ${\cal R}$, and keeping track of the term proportional to $(E_\mathrm{m}- E_0)$ one obtains 
\begin{eqnarray}\nonumber
	S \big( \mathrm{tr}\!_{{\cal L}\backslash{\cal R}} |\Phi \rangle\!\langle \Phi | \big) 
	&\leq& |\partial {\cal R}|
	\ln\!|{\cal R}|
	\big(\gamma 2 J 3^s + \eta + 4\xi \ln q \big)  
	\\ \nonumber
	&+& (E_\mathrm{m}- E_0) 
	(\ln\tau|{\cal R}|)^{1-\nu} \, 
	\frac{\gamma c_1 2^{\nu+3}}{\nu\xi}
	\\ \label{S cubic 2}
	&+& {\cal O}\big( |\partial {\cal R}| \big)\ .
\end{eqnarray}

\section{Conclusions} 

It is shown that ground states and low-energy states obey an area law for the entropy, provided two conditions hold: (i) the state has a sufficient decay of correlations, and (ii) the number of eigenstates with vanishing energy-density is not exponential in the volume of the system. 

\medskip A universal property for local hamiltonians is also here established. The energy fluctuations of eigenstates inside an arbitrary region can be observed by measuring the exterior and a superficial shell of the region. This extends to any pure state that can be written as a superposition of eigenstates with similar energy. 

\medskip Some thermodynamic quantities at finite temperature only depend on the density of states. Examples are: free energy, (global) entropy, heat capacity, etc. This paper establishes a relation between these thermodynamic quantities and ground-state entanglement.

\acknowledgments The author is very thankful to Ignacio Cirac and Guifre Vidal. This work has been financially supported by Caixa Manresa. Additional support has come from the Spanish MEC project TOQATA (FIS2008-00784) and QOIT (Consolider Ingenio 2010), ESF/MEC project FERMIX (FIS2007-29996-E), EU Integrated Project SCALA, EU STREP project NAMEQUAM, ERC Advanced Grant QUAGATUA.

\appendix 

\section{proofs}

\medskip\noindent{\bf Lemma 1. } Let $Q, \tilde{Q}$  be the operators defined in (\ref{Q}), (\ref{tilde Q}), then
\begin{equation}\label{lemma 2}
	\big\| Q- \tilde{Q} \big\| \leq |{\cal X}|^3\,  e^{-l}\ .
\end{equation}

\medskip\noindent{\em Proof } First, express $Q$ and $\tilde{Q}$ with a single integral, by using the identity
\[
	\int_{-\infty}^{e} \hspace{-3mm} d\omega \int\! \frac{dt}{2\pi}\, 
	e^{-\frac{\sigma t^2}{2}}\, e^{-i\omega t} =
	\int\! \frac{dt}{2\pi}\, \frac{e^{-\frac{\sigma t^2}{2}}}{0^+ -it}\, e^{-ie t}\ .
\]
Second, define the operators
\begin{eqnarray*}
	H_0 &=& H_{{\cal S}\cup {\cal S}'} - H_{\cal S}- H_{{\cal S}'}\ ,
	\\
	H_1 &=& H - H_{{\cal X}}- H_{\bar{\cal X}}\ ,	
\end{eqnarray*}
which respectively act on the regions ${\cal H}_0 , {\cal H}_1 \subset {\cal L}$. Note that $\mathrm{d}({\cal H}_0 , {\cal H}_1) =l-4$, $|{\cal H}_0| \leq |{\cal X}|$ , $|{\cal H}_1| \leq 5^s |{\cal X}|$, and
\begin{eqnarray*}
	&&\big \| e^{i H_{\cal X}t} e^{-iHt} 
	-e^{i H_{{\cal S}} t} e^{-i H_{{\bar{\cal X}} \cup {\cal S}} t} \big\|
	\\ &=& 
	\big\| e^{i (H-H_1)t} e^{-iHt} 
	-e^{i (H-H_1-H_0) t} e^{-i (H-H_0) t} \big\|\ .
\end{eqnarray*}
This, the triangular inequality, Lemma 2, and the Lieb-Robinson bound (\ref{LR}), give
\begin{eqnarray*}
	&& \big\| Q -\tilde{Q} \big\| 
	\\ &\leq& 
	2 \int_0^{t_0} \hspace{-2mm} dt\,
	\frac{1}{2\pi t}\,
	\big\| e^{i H_{\cal X}t} e^{-iHt} 
	-e^{i H_{\cal S} t} e^{-i H_{{\bar{\cal X}} \cup {\cal S}} t} \big\|
	\\ && +\ 
	2 \int_{t_0}^\infty \hspace{-3mm} dt\,
	\frac{e^{-\frac{\sigma t^2}{2}}}{2\pi t}\, 2
	\\ &\leq& 
	\frac{2 |{\cal X}|^3 J^2 5^{s}}{\pi}\,
	\int_0^{t_0} \hspace{-2mm} dt\,
	\frac{1}{t}
	\int_0^t \hspace{-2mm} dt_2 \int_0^{t_2} \hspace{-2mm} dt_1\,
	\frac{(vt_1)^{\lfloor l/2-2 \rfloor}}{\lfloor l/2-2\rfloor!} 
	\\ && +\ 
	\frac{4}{t_0 \sqrt{\sigma}}\, e^{-\frac{\sigma t_0^2}{2}}
	\\ &\leq& 
	\frac{|{\cal X}|^3}{\pi 5^{s} (l-1) }\,
	\frac{(v t_0)^{\lfloor l/2\rfloor}}{\lfloor l/2\rfloor !} 
	+ 
	\frac{4}{t_0 \sqrt{\sigma}}\, e^{-\frac{\sigma t_0^2}{2}}
\end{eqnarray*}
Putting $t_0= \lfloor l/2 \rfloor /(e^3 v)$ and using Stirling's approximation 
\[
	\frac{(vt_0)^{\lfloor l/2 \rfloor }}{\lfloor l/2 \rfloor !} \leq e^{\lfloor l/2 \rfloor \ln\frac{evt_0}{\lfloor l/2 \rfloor}}
	\leq e^{1 -l}\ .
\]
Putting $\sigma = 10^4 v^2/l \geq 2 l t_0^{-2}$ one obtains
(\ref{lemma 2}). \hfill $\Box$

\medskip\noindent{\bf Lemma 2. } Let $H,X,Y$ be hermitian matrices and $t>0$, then
\begin{eqnarray}\label{lemma 1}
	&& \big\| e^{i(H-X)t} e^{-iHt} - e^{i(H-X-Y)t} e^{-i(H-Y)t} \big\|
	\nonumber \\ &\leq& 
	\int_0^t \hspace{-2mm} dt_2 \int_0^{t_2} \hspace{-2mm} dt_1
	\big\| [X(t_1), Y] \big\| \ ,
\end{eqnarray}
where $X(t) = e^{iHt} X e^{-iH t}$.

\medskip\noindent{\em Proof } If $f(t)$ is a differentiable function with $f(0)=0$ then $f(t)= \int_0^t dt_1 f'(t_1)$. This implies the following two equalities. The following two inequalities are a consequence of the triangular inequality for the operator norm.
\begin{eqnarray*}
	&& \big\| e^{i(H-X-Y)t} e^{-i(H-Y)t} e^{iHt} e^{-i(H-X)t} - \mathbb{I} \big\|
	\\ &=& 
	\Big\| \int_0^t \hspace{-2mm} dt_2\, e^{i(H-X-Y)t_2}
	\big[ -iX e^{-i(H-Y)t_2} e^{iHt_2} 
	\\ && +\ 
	e^{-i(H-Y)t_2} e^{iHt_2}\, iX \big] e^{-i(H-X)t_2} \Big\|
	\\ &\leq& 
	\int_0^t \hspace{-2mm} dt_2\, \big\|-X + 
	e^{-i(H-Y)t_2} e^{iHt_2} X e^{-iHt_2} e^{i(H-Y)t_2}\big\|
	\\ &=& 
	\int_0^t \hspace{-2mm} dt_2\, \Big\| \int_0^{t_2} \hspace{-2mm} dt_1\, 
	e^{-i(H-Y)t_1} \big[Y, X(t_1)\big] e^{i(H-Y)t_1}\Big\|
	\\ &\leq& 
	\int_0^t \hspace{-2mm} dt_2 \int_0^{t_2} \hspace{-2mm} dt_1
	\big\| [X(t_1 ),Y] \big\| 
\end{eqnarray*}
\hfill $\Box$

\medskip\noindent {\bf Lemma 3. } The operator $\tilde{Q}$ defined in (\ref{tilde Q}) with $e_\mathrm{cut}= 2 J 3^s |\partial {\cal X}| + e_0 + 20v$, and the projector $P^\bot$ defined by (\ref{P}), satisfy
\begin{eqnarray}\label{tQ}
	&\langle \tilde{Q}\rangle& \geq\  \frac{1}{2} - 
	2 |{\cal X}|^3 e^{-l}\ ,
\\ \label{QP}
	&\langle P^\bot \tilde{Q} \rangle &
	\leq\   
	2 |{\cal X}|^3 e^{-l}
\ .
\end{eqnarray}

\medskip\noindent {\em Proof .} The positive operator
\begin{equation}
	M = \sum_n 
	\int_{-\infty}^{e_\mathrm{cut} -e_n}
	\hspace{-5mm} d\omega\ (2\pi \sigma)^{-1/2}\, 
	e^{-\frac{\omega^2}{2\sigma}}\, 
	|\psi_n \rangle\!\langle \psi_n|	
\end{equation}
allows for writing equality (\ref{step}) as 
\begin{equation}\label{QM}
	Q |\Psi_0\rangle = M|\Psi_0\rangle\ .
\end{equation}
The two projectors
\begin{equation}\label{Qpm}
	M_\pm = \sum_{n:\, e_n \leq e_\mathrm{cut} \pm \delta} 
	|\psi_n\rangle\!\langle \psi_n|\ ,
\end{equation}
with $\delta= 20 v$, satisfy 
\begin{equation}\label{Q- Q Q+}
	M_- -e^{-l} \mathbb{I}\ \leq\ M\ 
	\leq\ M_+ +e^{-l} \mathbb{I}\ ,
\end{equation}
where we have used that $e^{-\frac{\delta^2}{2\sigma}} \leq e^{-l}$. The positivity of $M$ and the second inequality in (\ref{Q- Q Q+}) imply
\begin{equation}\label{M^2}
	M^2 \leq (1+2 e^{-l}) M_+ + e^{-2l}\ .
\end{equation}
A worst-case estimation gives
\begin{equation}\label{jbp}
	\langle H_{\cal X} \rangle \geq \langle M_- \rangle e_0 + 
	\langle \mathbb{I} - M_-\rangle (e_\mathrm{cut} -\delta)\ .
\end{equation}
Performing the assignation $e_\mathrm{cut}= 2 J 3^s |\partial {\cal X}| + e_0 +\delta$ in (\ref{jbp}) and using bound (\ref{e bound}) one obtains $\langle M_-\rangle \geq 1/2$. The combinations of (\ref{Q tildeQ}), (\ref{QM}) and (\ref{Q- Q Q+}) gives (\ref{tQ}).

\medskip Using Lemma 1 and (\ref{QM}), the Cauchy-Schwarz inequality, bound (\ref{M^2}), and the definition of $M_+$ and $P^\bot$, one obtains respectively the following chain of inequalities:
\begin{eqnarray}
	\nonumber
	&& \langle P^\bot \tilde{Q} \rangle 
	\ \leq\  
	\langle P^\bot M \rangle + |{\cal X}|^3 e^{-l}
	\\ \nonumber &\leq& 
	\langle P^\bot \rangle^{1/2}  
	\langle P^\bot M^2 P^\bot \rangle^{1/2} 
	+ |{\cal X}|^3 e^{-l}
	\\ \nonumber &\leq& 
	\left[ (1+2 e^{-l}) \langle P^\bot M_+ P^\bot \rangle 
	+ e^{-2l} \right]^{1/2} + |{\cal X}|^3 e^{-l}
	\\ \label{QPp} &\leq& 
	2 |{\cal X}|^3 e^{-l}
\ ,
\end{eqnarray}
which is (\ref{QP}). \hfill $\Box$

\section{calculation of the entropy}

\subsection{Entropy of an arbitrary region}

Consider the probability distribution defined by
\begin{eqnarray}\label{prob}
	p_k &=& 
	\frac{1- \theta(l_0)}{\Theta(l_0)}
	\quad \mathrm{for}
	\quad 1 \leq k \leq \Theta(l_0)\ ,
	\\ \nonumber
	p_k &=& \frac{\theta(l-1) - \theta(l)}{\Theta(l)-\Theta(l -1)} 
	\quad \mathrm{for} 
	\quad \Theta(l-1)+1 \leq k \leq \Theta(l)\ ,
\end{eqnarray}
for every integer $l \geq l_0 = 2\xi\ln |{\cal R}|$, and
\begin{eqnarray}
	\label{eta l}
	\theta(l) &=& \frac{4\, c_1}{(l- \xi\ln |{\cal R}|)^{\nu}}\ ,
	\\ \nonumber
	\ln \Theta(l) &=& |\partial {\cal R}| 2s (5l)^{s-1}
	\big( \gamma 2 J 3^s + \eta \big) \ln\!\big[\tau |{\cal R}| (5l)^s\big]
	\\ \nonumber && 
	+\ |\partial {\cal R}| (5l)^s \ln q 
	+{\cal O} \big( \ln\! |{\cal R}| \big)\ .
\end{eqnarray}
This distribution is uniform in blocks of the maximum size that constraints (\ref{constr}) allow. Then, it is the distribution satisfying (\ref{constr}) with maximum entropy. The upper-bound on the entropy of $p_k$ gets simplified by using the substitutions $\Theta(l)- \Theta(l-1) \leq \Theta(l)$ and
\begin{equation}\label{truc}
	\theta(l-1)- \theta(l) \leq 
	\frac{c_1 2^{\nu+3}}{(l-\xi\ln|{\cal R}|)^{\nu +1}}\ .
\end{equation}
Using this, one obtains
\begin{eqnarray}\nonumber
	-\sum_k p_k \ln p_k &\leq& |\partial {\cal R}|
	(10\xi \ln\!|{\cal R}|)^s
	\left[\frac{s}{\xi} \big(\gamma J 3^s + \eta \big)+ \ln q\right]  
	\\ &&+\  {\cal O}\big[ |\partial {\cal R}|
	(\ln\!|{\cal R}|)^{s-1} \big]\ .
\end{eqnarray}

\subsection{Entropy of a cubic region}

Consider the case where the chosen region is an hypercube ${\cal R} = \{x\in{\cal L}: 0\leq x_i \leq L\}$. It is easy to calculate
\begin{eqnarray*}
	|{\cal R}| &=& L^s\ ,
	\\
	|\partial {\cal R}| &=& 2 s L^{s-1}\ .
\end{eqnarray*}
Following definitions (\ref{def X}, \ref{def dX}) one obtains
\begin{eqnarray*}
	|{\cal X}| &=& |{\cal R}| (1+4l/L)^s\ ,
	\\
	|\partial {\cal X}| &=& |\partial {\cal R}| (1+4l/L)^{s-1}\ ,
	\\
	|{\cal X}\backslash {\cal R}| &\leq& |\partial{\cal X}| 2l\ .
\end{eqnarray*}
Consider the probability distribution (\ref{prob}) with $\theta(l)$ given in (\ref{eta l}) but $\Theta(l)$ defined as
\begin{eqnarray}\nonumber
	\ln \Theta(l) &=& |\partial {\cal R}| (1+4l/L)^{s-1}
	\big( \gamma 2 J 3^s + \eta \big) \ln\!\big[\tau |{\cal R}| (1+4l/L)^s \big]  
	\\ \nonumber && 
	+\ |\partial {\cal R}| (1+4l/L)^{s-1} 2l\ln q 
	+ {\cal O}( \ln\! |{\cal R}| )\ .
\end{eqnarray}
Using the same tricks as above one obtains the following upper-bound for the entropy of $p_k$,
\[
	-\sum p_k \ln p_k \leq |\partial {\cal R}|
	\ln\!|{\cal R}|
	\big(\gamma 2J 3^s +\eta +4\xi \ln q \big)  
	+\ {\cal O} \big( |\partial {\cal R}| \big)\ .
\]

\section{The Lieb-Robinson bound}

Let $X,Y$ be two operators with support on the regions ${\cal X}, {\cal Y}$ respectively, and $L= \mathrm{d}({\cal X}, {\cal Y})$. Let $Z$ be an arbitrary operator and $F(t) =[X(t),Z]$, where $X(t)= e^{iHt} X e^{-iHt}$ and $H$ is the hamiltonian (\ref{H}). Using the Jacobi identity $[[X,Y],Z] + [[Y,Z],X] + [[Z,X],Y] =0$ twice one obtains
\begin{eqnarray}
	\partial_t F(t) &=& i[[H,X(t)],Z] 
\\ \nonumber
	&=& -i[F(t),H] -i[[Z,\mbox{$\sum_{x\in {\cal Z}}$} K_x],X(t)]
\\ \nonumber
	&=& i[A, F(t)] +i \mbox{$ \sum_{x\in {\cal Z} }$} [Z,[X(t),K_x]]\ ,
\end{eqnarray}
where ${\cal Z} =\{x: [K_x,Z]\neq 0\}$ and $A= \sum_{x \in {\cal L} \backslash {\cal Z}} K_x$. The above is equivalent to
\[
	\partial_t \left( e^{-iAt} F(t)\, e^{iAt}\right)
	= i\sum_{x\in {\cal Z}} e^{-iAt} [Z,[X(t),K_x]]\, e^{iAt}\ ,
\]
which can be integrated 
\begin{eqnarray}
	&& e^{-iAt} F(t)\, e^{iAt}
	\\ \nonumber
	&=& F(0) +
	i\sum_{x\in {\cal Z}} \int_0^t \hspace{-2mm} dt_1\, e^{-iA t_1} [Z,[X(t_1),K_x]] e^{iA t_1}\ .
\end{eqnarray}
The triangular inequality for the operator norm gives
\begin{eqnarray}\label{Z}
	&& \| [X(t),Z] \|
	\\ \nonumber
	&\leq& \| [X(0),Z] \| +
	2 \|Z\|\sum_{x\in {\cal Z}} \int_0^t \hspace{-2mm} dt_1\, \| [X(t_1),K_x]] \| \ .
\end{eqnarray}
Define $g_x (t) = \|[X(t), K_x]\|$ and use (\ref{Z}) with $Z=K_x$ to obtain
\[
	g_x (t) \leq g_x (0) + 2J\hspace{-4mm} \sum_{x':\, \mathrm{d}(x,x')\leq 2 } 
	\int_0^t \hspace{-2mm} dt_1\, g_{x'} (t_1)\ .
\]
If $r=\mathrm{d}(x,{\cal X}) \geq 2$ then $g_x (0)=0$. The above can be iterated $n= \lfloor (r-1)/2 \rfloor$ times
\[
	g_x (t) \leq v^n \hspace{-3mm} \max_{x':\, \mathrm{d}(x,x')\leq 2n } 
	\int_0^t \hspace{-2mm} dt_n \int_0^{t_n} \hspace{-2mm} dt_{n-1} \cdots \int_0^{t_2} \hspace{-2mm} dt_1\, g_{x'} (t_1)\ ,
\]
where $v= 2J 5^s$ and $|\{x':\, \mathrm{d}(x,x')\leq 2\}| =5^s$. For any $x'$ the bound $g_{x'} (0) \leq 2 J \|X\|$ holds, then
\begin{equation}\label{g}
	g_x (t) \leq 2J \|X\|\, \frac{(vt)^{\lfloor (r-1)/2 \rfloor }}{\lfloor (r-1)/2 \rfloor !} \ .
\end{equation}
Note that $|\{ x: [K_x,X] \neq 0 \}| \leq 5^s |{\cal X}|$. This and the bound (\ref{g}) can be substituted in (\ref{Z}) with $Z=Y$, giving
\[
	\|[X(t),Y]\| \leq 2 |{\cal X}|\, \|X\| \|Y\|\, \frac{(vt)^{\lfloor L/2 \rfloor}}{\lfloor L/2 \rfloor !} \ .
\]
This is the Lieb-Robinson bound.

\section{Simpler proof for the area law}

Let us obtain an upper-bound for the entropy
\begin{equation}\label{Sm}
	S= -\sum_n \mu_n \ln \mu_n\ ,
\end{equation}
of any probability distribution $\mu_n$ subjected to the constraints 
\begin{eqnarray}\label{con1}
	n &\leq& c_2 (\tau |{\cal R}|)^{\gamma (e_n-e_0) +\eta |\partial {\cal R}|}\ ,
\\ \label{con2}
	\sum_n \mu_n e_n &\leq& e_0 + J3^s |\partial {\cal R}|\ .
\end{eqnarray}
The original problem has a finite range for $n$, but relaxing this fact cannot diminish the largest entropy (\ref{Sm}). Isolating $e_n$ from (\ref{con1}) gives
\[
	e_n\geq \frac{\ln(n/c_2)}{\gamma \ln (\tau |{\cal R}|)} 
	-\frac{\eta}{\gamma} |{\cal R}| +e_0\ ,
\]
which substituted in (\ref{con2}) gives
\begin{equation}\label{con3}
	\sum_n \mu_n \ln n \leq 
	|\partial {\cal R}| \ln (\tau |{\cal R}|)\, (J3^s\gamma +\eta)
	+ \ln c_2 =c_3\ .
\end{equation}
The new constant $c_3$ is define through this equality. For any $a\geq 0$, the entropy (\ref{Sm}) can be upper bounded as
\begin{equation}\label{smax}
 S \leq\ \max_{\{\mu_n \}} \Big[
 -\sum_n \mu_n \ln\mu_n 
 + a \Big( c_3 - \sum_n \mu_n \ln n \Big)
 \Big]\ .
\end{equation}
This upper-bound holds because the second term between the square brackets is always positive. In what follows, we perform the maximization (\ref{smax}) with no constrains over $\{\mu_n \}$: normalization, constrains (\ref{con1}) and (\ref{con2}) are ignored. This relaxation cannot decrease the right-hand side of (\ref{smax}). Deriving the right-hand side of (\ref{smax}) with respect to $\mu_n$ and equating to zero gives
\[
	-\ln \mu_n -1 -a\ln n =0 \ ,
\]
which implies $\mu_n = n^{-a}/e$. Substituting this into (\ref{smax}) gives 
\[
	S\ \leq\ \sum_n n^{-a}/e + a c_3\ . 
\]	
Choosing $a=2$ and replacing $c_3$ gives
\[
	S \leq\  
	2J3^s \gamma\, |\partial {\cal R}| \ln (\tau |{\cal R}|)
	+2\eta |\partial {\cal R}| + 2 \ln c_2 + \pi^2/(6e)\ .
\]

\end{document}